\documentclass[sigconf]{acmart}

\AtBeginDocument{%
  }

\setcopyright{acmlicensed}
\copyrightyear{2018}
\acmYear{2018}
\acmDOI{XXXXXXX.XXXXXXX}

\acmConference[EASE 2026]{The 30th International Conference on Evaluation and Assessment in Software Engineering}{9–12 June, 2026}{Glasgow, Scotland, United Kingdom}

\acmISBN{978-1-4503-XXXX-X/18/06}

\usepackage{bm}

\usepackage{xcolor}

\usepackage{array}

\usepackage{listings}
\newcommand*{\mytt}{\fontfamily{lmtt}\selectfont}
\definecolor{javared}{rgb}{0.6,0,0}
\definecolor{javagreen}{rgb}{0.25,0.5,0.35}
\definecolor{javapurple}{rgb}{0.5,0,0.35}
\definecolor{javadocblue}{rgb}{0.25,0.35,0.75}
\lstdefinelanguage{xml}{
    basicstyle=\mytt\footnotesize,
    morestring=[b]",
    stringstyle=\color{javared},
    morecomment=[s]{<!--}{-->}
}
\lstdefinelanguage{csv}{
  basicstyle=\mytt\footnotesize,
  columns=fullflexible,
  breaklines=true,
  postbreak=\mbox{\textcolor{lightgray}{$\hookrightarrow$}\space}
}

\usepackage{pifont}

\usepackage{booktabs}
\usepackage{multirow}
\usepackage[flushleft]{threeparttable}

\usepackage{lipsum}

\usepackage{tcolorbox}
\tcbset{
    top=1pt,
    left=1pt,
    right=1pt,
    bottom=1pt,
    toprule=1pt,
    titlerule=0pt,
    bottomrule=1pt,
    leftrule=1pt,
    rightrule=1pt
}

\newcounter{finding}
\newcommand*{\newfinding}{Finding \stepcounter{finding}\thefinding:~}

\usepackage{marvosym}
\newcommand*{\downredarrow}{\color{red!80!cyan}{\rotatebox[origin=c]{-90}{\MVRightarrow}}}
\newcommand*{\upgreenarrow}{\color{green!60!teal}{\rotatebox[origin=c]{90}{\MVRightarrow}}}

\usepackage[shortlabels]{enumitem}
\setlist[itemize]{align=parleft,left=0pt..1em}
\setlist[enumerate]{align=parleft,left=0pt..1.5em}

\definecolor{tiffany}{HTML}{0abab5}

\begin{document}

\title{Towards Understanding Android APIs: Official Lists, Vendor Customizations, and Real-World Usage}

\author{Sinan Wang}
\orcid{0000-0002-8557-4681}
\authornote{Sinan Wang and Yepang Liu are affiliated with the Research Institute of Trustworthy Autonomous Systems (RITAS) at Southern University of Science and Technology.}
\affiliation{%
  \institution{Southern University of Science and Technology}
  \city{Shenzhen}
  \state{Guangdong}
  \country{China}}
\email{wangsn@mail.sustech.edu.cn}

\author{Qi Zhang}
\orcid{0009-0001-0319-0191}
\affiliation{%
  \institution{Hong Kong University of Science and Technology}
  \city{Hong Kong SAR}
  \country{China}}
\email{qzhangdh@cse.ust.hk}

\author{Jiacheng Li}
\orcid{0009-0009-9472-2357}
\affiliation{%
  \institution{University of Maryland}
  \city{College Park}
  \state{MD}
  \country{USA}}
\email{jiacli@umd.edu}

\author{Lili Wei}
\orcid{0000-0002-2428-4111}
\affiliation{%
 \institution{McGill University}
 \city{Montr\'eal}
 \state{QC}
 \country{Canada}}
\email{lili.wei@mcgill.ca}

\author{Yida Tao}
\orcid{0000-0001-6866-336X}
\affiliation{%
  \institution{Southern University of Science and Technology}
  \city{Shenzhen}
  \state{Guangdong}
  \country{China}}
\email{taoyd@sustech.edu.cn}

\author{Yepang Liu}
\orcid{0000-0001-8147-8126}
\authornotemark[1]
\authornote{Yepang Liu is the corresponding author.}
\affiliation{%
  \institution{Southern University of Science and Technology}
  \city{Shenzhen}
  \state{Guangdong}
  \country{China}}
\email{liuyp1@sustech.edu.cn}

\renewcommand{\shortauthors}{Wang et al.}

\begin{abstract}

Android apps are built on APIs that abstract core Android system functionalities.
These APIs are officially documented in multiple files distributed with the Android source code or SDK, which we collectively refer to as \textbf{Android API Lists (AALs)}.
Prior Android research has relied on specific AALs, often treating them as interchangeable ground truth.
However, recent studies suggest that different AALs can lead to substantially different research outcomes, raising concerns about the validity and reproducibility of Android API-based analyses.
To address this issue, we present the first in-depth empirical study of four official AALs that are widely used in prior work.
We systematically characterize their contents and analyze their evolution across Android releases.
We then perform a fine-grained comparison of the APIs recorded in each AAL to uncover their underlying API inclusion policies and inconsistencies.
To assess the practical impact of these differences, we further examine API availability on nine Android devices, including both stock Android and vendor-customized systems.
Finally, we analyze API usage in 17,759 real-world Android apps (including open-source apps, commercial apps, and malware) to quantify how the choice of AAL affects empirical Android research.
Our results reveal that official AALs are neither stable nor mutually consistent, and that discrepancies among them can substantially influence research conclusions.
We also observe that vendor-customized APIs are actively used by normal apps, yet remain largely overlooked by existing studies.
Based on these findings, we discuss their implications for Android API-based research and provide actionable suggestions to help researchers select and interpret AALs more reliably.

\end{abstract}





\maketitle

\section{Introduction}
\label{sec:introduction}


The Android framework provides a set of APIs that abstract the core Android system functionalities and hardware features, thereby supporting the functioning of running apps.
These Android APIs ease app development, but they also bring API compatibility issues \cite{ficfinder-ase16}.
It has been shown that those change-prone Android APIs can trigger more developers' discussions \cite{linares-icpc14} and even threat to the apps' ratings~\cite{linares-fse13}.
At the user end, one may find a particular app only works on specific Android versions or devices, causing poor user experience.
Android SDK \cite{web:android/sdk-manager} provides facilities to reduce developers' maintaining effort in this situation.
For example, AndroidX~\cite{li2023towards} helps maintain an API's backward compatibility. 
\textsc{Lint} \cite{web:android/lint} is an another SDK tool, which provides a rule-based code smell scanning service.
It can detect certain API compatibility issues \cite{confdroid-ase21,aper-icse2022} in the application source code.
However, \textsc{Lint}'s performance have been shown to be weak \cite{ficfinder-ase16,ficfinder-tse2018,cider-ase18,aper-icse2022}.


These compatibility issues attract researchers' attention.
Wei et al. \cite{ficfinder-ase16}  developed \textsc{FicFinder}, a tool for identifying compatibility issues via pairs of APIs and their respective issue-triggering contexts (i.e., API-context pairs).
However, \textsc{FicFinder} depends on a set of manually written API-context pairs, limiting its practical usability in the evolving Android ecosystem.
Later on, static compatibility issue detectors mostly involve a separate module for systematic API harvesting.
For example, \textsc{CiD} \cite{cid-issta18} has an API Lifecycle Modeling module.
It reads the source code of the Android framework and extracts APIs' lifetimes for API-based issue detection.
\textsc{IctApiFinder}~\cite{ictapifinder-ase18}, another compatibility issue detector, collects all APIs in the \texttt{android.jar} files bundled in Android SDK.
Mahmud et al. proposed ACID \cite{acid-saner21}, which leverages Android API differences report \cite{web:android/api-diff-report} for detecting two kinds of API compatibility issues.
As pointed out by Liu et al. \cite{liu2022automatically}, tools that have systematic API harvesting methods can detect significantly more issues than those that do not.
They also found that tools using different sources of API knowledge output few overlapped results.


The research work mentioned so far all rely on specific Android API knowledge.
The accuracy of the research outcomes significantly depend on the utilization of carefully curated API lists.
We call such curated API files \textit{\underline{A}ndroid \underline{A}PI \underline{L}ists} (AALs for short).
Despite the pervasive uses of AALs in previous work, little is known about whether they faithfully represent the Android API knowledge.
In fact, we revealed that, even in the same Android version, different AALs may record distinct API knowledge, and the recorded APIs in an AAL may not reflect the actual APIs in a running Android device.
For example, on Android 13, four Java methods were removed from the AAL \texttt{current.txt}, though they retained in the Android framework and could be used by normal apps.
This AAL is utilized by the state-of-the-art API compatibility issue detector, \textsc{PSDroid}.
Incomplete API inclusion brings false negatives to \textsc{PSDroid}'s detection results.
In other words, incomplete or incorrect AALs pose substantial threats to various research outcomes.


\textbf{Our contribution:}
A better understanding of the Android API lists helps Android researchers and practitioners make more informed decisions in the future.
To this end, we conducted the first in-depth study about four official AALs under six Android release versions.
Specifically, we studied how these AALs include APIs from Android system, how they change along with the evolution of Android platform, and how they are different from each other.
We further explore the API existence on nine real Android devices and the API usage situations in 17,759 real Android apps.
Our study demystifies the API instability and inconsistencies among the four official AALs, as well as their impact on the Android ecosystem.
Our research reveals the following key findings:

\begin{enumerate}
    \item \textbf{AALs exhibit substantial inconsistencies}, with only a few (\textasciitilde10\%) APIs shared among all investigated AALs.
    \item \textbf{AALs may include considerable amount of synthesized APIs}, which are presented in some AALs but absent from the Android source code or real Android devices.
    \item \textbf{No single AAL completely records all APIs},  especially for non-SDK interfaces and vendor-customized APIs.
    \item \textbf{AALs' API inclusion policies are unstable}, which will change during platform evolution.
\end{enumerate}

\noindent
We also highlight research opportunities in the context of Android APIs, with an emphasis on the non-SDK APIs and vendor-customized APIs.
To support future work, we release all our research artifacts on \url{https://github.com/sqlab-sustech/AAL-artifacts}, including AAL dataset, source code of tools, and experiment results.

\section{Android API Lists}
\label{sec:api-lists}

\subsection{Research Scope}
\label{ssec:research-scope}

\begin{table*}[t!]
\caption{Tools or studies related to Android APIs, and how they obtain API knowledge}
\label{tab:existing-tools}
\resizebox{\textwidth}{!}{
\newcolumntype{C}[1]{>{\centering\arraybackslash}p{#1}}
\begin{tabular}{@{}ccllC{1cm}@{}}
\toprule
\textbf{Tool/Study}                               & \textbf{Year}  & \multicolumn{1}{c}{\textbf{Research Topic}} & \multicolumn{1}{c}{\textbf{Depending API Knowledge}} & \textbf{AAL} \\ \midrule
\textsc{PSDroid}~\cite{psdroid-icse2023}          & 2023                     & API compatibility issue detection           & \texttt{current.txt} file in Android source code & TXT                               \\
ACID~\cite{acid-tse23}                            & 2023                      & API compatibility issue detection           & API differences report & -                                                       \\
\textsc{APIMatchmaker}~\cite{apimatchmaker-tse22} & 2022                      & API recommendation                          & parse Android source code for APIs with lifetime & -                    \\
Yang et al.~\cite{yang-icse2022}                  & 2022                     & empirical study of non-SDK interfaces       & \texttt{hiddenapi-flags.csv} file from Android source code & CSV                \\
A3~\cite{a3-tse2020}                              & 2020                      & API migration                               & API documentation, Android source code & -                                   \\
RAPID~\cite{rapid-icse2020}                       & 2020                     & API compatibility issue detection           & \texttt{api-versions.xml} file from Android SDK & XML                        \\
\textsc{AppEvolve}~\cite{appevolve-issta19}       & 2019                    & API migration                               & API differences report & -                                               \\
\textsc{IctApiFinder}~\cite{ictapifinder-ase18}   & 2018                      & API compatibility issue detection           & \texttt{android.jar} file from Android SDK & JAR                            \\
\textsc{CiD}~\cite{cid-issta18}                   & 2018                    & API compatibility issue detection           & parse Android source code for APIs with lifetime & -               \\
CDA~\cite{cda-msr18}                              & 2018                      & empirical study of deprecated APIs          & parse Android source code for all deprecated APIs & -                  \\
Li et al.~\cite{li-icsme16}                       & 2016                    & empirical study of inaccessible APIs        & parse Android source code for all hidden/internal APIs & -            \\
McDonnell et al.~\cite{mcdonnell-icsm2013}        & 2013                     & empirical study of API evolution            & \texttt{api-versions.xml} file from Android SDK & XML                       \\ \bottomrule
\end{tabular}

}
\end{table*}

Table~\ref{tab:existing-tools} lists recent research work that rely on specific Android API knowledge.
Among the listed papers, five of them directly utilize the bundled Android API lists from either Android source code or Android SDK, which consist of the four bundled AALs involved in our study.
To ease presentation, we denote each of them according to their file extensions, that is, JAR for \texttt{android.jar}, XML for \texttt{api-versions.xml}, TXT for \texttt{current.txt} and CSV for \texttt{hiddenapi-flags.csv}, respectively.

We do not investigate APIs from the web documentation due to potential inaccuracies, as highlighted in \cite{a3-tse2020}.
We also do not consider Android source code as a separate AALs.
Instead, we place confidence in it as a dependable reference when investigating the correctness of the investigated AALs.

\subsection{API Definitions}
\label{ssec:api-definition}

Android framework APIs are written in Java, which means they are either \textit{fields} or \textit{methods} of some Java \textit{classes}.
Hence we consider APIs to be these three types of Java language constructs.
To uniquely identify a Java API, we follow the practice of Java reflection \cite{li2019understanding} on how it retrieves classes and their members.
Specifically, we denote a class by its \textit{full\_class\_name} (a.k.a. fully-qualified name), which consists of the class's package and its short name.

A field in the class is denoted by a pair of character strings:
$$\langle \mathit{full\_class\_name},\mathit{field\_name}\rangle$$

A method can be uniquely identified by a 3-tuple of strings:
$$\langle \mathit{full\_class\_name},\mathit{method\_name},\mathit{parameter\_list}\rangle$$
while the last term is a (possibly empty) list of strings of the method's formal parameter types.

With such notation, constructors can be denoted as methods that have a special name "\texttt{\textless{init}\textgreater}" \cite{web:java/jvm2-2.9}.
It is worth noting that our notation sufficiently includes nested classes and their members.
Also notice that callbacks are special methods whose control are inverted to the application framework \cite{cider-ase18}, thus our notation can cover them without additional adaptation.

\subsection{The JAR List}

\begin{figure}[h]
\begin{lstlisting}[language=java]
package android.app;

import android.content.ContextWrapper;

public class Service extends ContextWrapper {
  public static final int START_STICKY = 1;
  // other API fields...

  @Deprecated
  public final void stopForeground(boolean removeNotification) {
      throw new RuntimeException("Stub!");
  }
  public final void stopSelf() {
      throw new RuntimeException("Stub!");
  }
  // other API methods...
}
\end{lstlisting}
    \caption{Decompiled code snippet in the \texttt{android.jar} file}
    \label{fig:android-jar}
\end{figure}%

The \texttt{android.jar} file is an essential part of the Android SDK.
It can be downloaded by the SDK Manager tool \cite{web:android/sdk-manager}, or built through the \texttt{sdk} target from the Android source code.
The file contains a collection of compiled Java classes and resources that define the Android application framework at a specific API-level.
By linking this file, developers gain access to the Android framework classes, fields, and methods.
This allows them to write Java code that utilizes the platform's features, such as telephony or locationing services.
Notably, this file does not contain the actual implementation of API methods.
Their actual implementations reside in the running devices. 
Figure \ref{fig:android-jar} shows the class \texttt{Service} in \texttt{android.jar} at API-level 33, which contains one integer constant field, and two method stubs.

\subsubsection{API Extraction}
We used Soot \cite{lam2011soot} to load all classes in an \texttt{android.jar} file.
Then, for each successfully loaded class, we iterated over its field and method lists to obtain all the declared APIs.
We also extracted classes that have no declared members, such as the \texttt{Serializable} interface.
It is worth mentioning that, we do not distinguish abstract classes, interfaces, enum classes, or annotation declarations.
All of them can be regarded as some special forms of Java classes, while their conceptual differences do not affect the results of our study.

\subsection{The XML List}

\begin{figure}[h]
\begin{lstlisting}[language=xml]
<class name="android/app/Service" since="1">
    <extends name="android/content/ContextWrapper"/>

    <method name="setForeground(Z)V" removed="11"/>
    <method name="stopForeground(Z)V" since="5" deprecated="33"/>
    <method name="stopSelf()V"/>
    <!-- other API methods... -->

    <field name="START_STICKY" since="5"/>
    <!-- other API fields... -->
</class>
\end{lstlisting}
    \caption{A class in the \texttt{api-versions.xml} file}
    \label{fig:api-versions-xml}
\end{figure}

The \texttt{api-versions.xml} file is also bundled in the Android SDK.
It is generated along with the \texttt{android.jar} file when building SDK.
By referencing this file, one can determine the minimum and maximum API-levels of an API.
This information is important for ensuring their apps are compatible with the desired range of Android versions.
\textsc{Lint} leverages this file to build a database, which supports quick lookup for the status of an API (e.g., available, deprecated) at certain API-levels.
Figure \ref{fig:api-versions-xml} shows a snippet of this file, which displays the same API information as Figure \ref{fig:android-jar}.
However, it provides more information about the API's lifetime, such as when the API was deprecated.
Notably, \textbf{this file also contains removed APIs}, indicated by the \texttt{removed} attribute on their XML elements.
This enables \textsc{Lint} to determine an API's lifetime.

\subsubsection{API Extraction}
We directly parsed the \texttt{api-versions.xml} file to extract all the declared APIs.
In particular, we dropped out those elements having a \texttt{removed} attribute, as we are interested in the available APIs at a specific API-level.

\subsection{The TXT List}
\label{ssec:txt}

\begin{figure}[h]
\begin{lstlisting}[language=java]
package android.content {
  public abstract class Context {
    // constructor
    ctor public Context();

    // API methods...
    method public final String getString(int,java.lang.Object...);
    method public final <T> T getSystemService(Class<T>);
    method public abstract String getSystemServiceName(Class<?>);

    // API fields...
    field public static final int MODE_WORLD_READABLE = 1; // 0x1
    field public static final String WINDOW_SERVICE = "window";
  }
}
\end{lstlisting}
    \caption{A simplified class entry in the \texttt{current.txt} file}
    \label{fig:current-txt}
\end{figure}

The TXT list refers to the \texttt{current.txt} files in Android \cite{web:android/current-txt}.
It is automatically updated once APIs are changed and the \texttt{update-api} build target is triggered \cite{web:stackoverflow/update-api}.
\textsc{PSDroid} \cite{psdroid-icse2023} utilizes this file to build an API lifetime database, which records when an API was being added to and deprecated from the Android framework.

A major difference between the TXT list and the other AALs is that \textbf{this file records the source-level API signatures}.
This format retains front-end language features that will be erased in bytecode, for examples,  generic types or variable arguments (a.k.a. vararg).
Source-level signatures help developers better understand the underlying semantics of an API.
However, the signatures can be different from that being referenced in the Android bytecode.
Figure \ref{fig:current-txt} shows an example class entry in the \texttt{current.txt} file at API-level 33.
The file is written in a special context-free grammar, in which each class has multiple sections: its constructors, methods, fields, and so on.
In Figure \ref{fig:current-txt}, the method \texttt{getString} has a vararg parameter \texttt{Object...}, which will be converted into \texttt{Object[]} in compilation.
The next two methods have a type argument \texttt{T} and an unbounded wildcard generic type parameter, respectively.
After compilation, the generic types are replaced by \texttt{Object}, and the type parameters are erased.
Then their method declarations become:

\begin{lstlisting}[frame=none,basicstyle=\mytt\small]
            String getString(int, Object[])
            Object getSystemService(Class)
            String getSystemServiceName(Class)
\end{lstlisting}

\noindent \textsc{PSDroid} employs such conversion to its extracted APIs, thus their database can be utilized to analyze API usages in Android bytecode.

\subsubsection{API Extraction}

\begin{figure}[t]

\begin{gather}
\frac{\textbf{\texttt{package}}\ \textit{Id}\ \textbf{\texttt{\{}}}
     {\bm{\phi}\ {\longrightarrow}_t\ \bm{\phi}[t \leadsto \textit{Id}]}
\notag \\
\tag*{(rule 1, package definition)}
\end{gather}

\begin{gather}
\frac{\textbf{\texttt{class}}\ \textit{Id}[\textbf{.}\textit{Id}]^*\ [(\textbf{\texttt{extends}}\ |\ \textbf{\texttt{implements}})\ \mathit{Type}]\ \textbf{\texttt{\{}}}{\bm{\theta}\ {\longrightarrow}_t\ \bm{\theta}[t \leadsto \textit{Id}\ [\textbf{.}\textit{Id}]^*]\ \ \ \ \bm{\textbf{C}}\ {\longrightarrow}_t\  \bm{\textbf{C}} \cup \{\langle\bm{\phi}[t]\textbf{.}\bm{\theta}[t]\rangle\}}
\notag \\
\tag*{(rule 2, class definition)}
\end{gather}

\begin{gather}
\frac{\textbf{\texttt{field}}\ \mathit{Type}\ \textit{Id}\ \mathit{[\textbf{=}\ Value]}\textbf{\texttt{;}}}{\bm{\textbf{F}}\ {\longrightarrow}_t\  \bm{\textbf{F}} \cup \{\langle\bm{\phi}[t]\textbf{.}\bm{\theta}[t],\ \textit{Id}\rangle\}}
\notag \\
\tag*{(rule 3, field definition)}
\end{gather}

\begin{gather}
\frac{\textbf{\texttt{method}}\ \mathit{[GenericTypeList]}\ \mathit{Type}\ \textit{Id}\ \textbf{\texttt{(}}[\mathit{TypeList}]\textbf{\texttt{)}}\textbf{\texttt{;}}}{\bm{\textbf{M}}\ {\longrightarrow}_t\  \bm{\textbf{M}} \cup \{\langle\bm{\phi}[t]\textbf{.}\bm{\theta}[t],\ \textit{Id},\ \mathit{TypeList}\rangle\} \ \ \ \ \bm{\Omega} \ {\longrightarrow}_t\  \varnothing}
\notag \\
\tag*{(rule 4, method declaration)}
\end{gather}

\begin{gather}
\frac{\mathit{Param}\ \textbf{\texttt{extends}}\ \mathit{Type}}{\bm{\Omega}\ {\longrightarrow}_t\ \bm{\Omega} \cup \{\langle\mathit{Param}\mapsto\mathit{Type}\rangle\}}
\notag \\
\tag*{(rule 5.1, bounded generic type)} 
\end{gather}

\begin{gather}
\frac{\mathit{Param}}{\bm{\Omega}\ {\longrightarrow}_t\ \bm{\Omega} \cup \{\langle\mathit{Param}\mapsto\texttt{"java.lang.Object"}\rangle\}}
\notag \\
\tag*{(rule 5.2, unbounded generic type)}
\end{gather}

\begin{gather}
\frac{\textit{Id}[\textbf{.}\textit{Id'}] ^* \ \ \ \ \ [\textbf{.}\textit{Id'}] ^* = \epsilon}{(\textbf{if}\ \exists.\Omega[id] \ \textbf{then}\ \Omega[\textit{Id}]\ \textbf{else}\ \textit{Id}) \leadsto \textit{Id}}
\notag \\
\tag*{(rule 6, generic type name)} 
\end{gather}

\caption{Parsing the TXT list file}
\label{txt:semantics}
\end{figure}

The TXT list is written in a special grammar that involves many source-level language features.
Following the API definitions in Section \ref{ssec:api-definition}, we implemented a parser to collect three sets of APIs: classes $\textbf{C}$, fields $\textbf{F}$, and methods $\textbf{M}$.
The parser also manipulates three state variables: the current package name $\bm{\phi}[t]$, the current class name $\bm{\theta}$[$t$], and a mapping of generic type bound $\bm{\Omega} \subset \mathit{\{\langle GenericType\ \mapsto\ Type\rangle^*\}}$.

The parser is shown in Figure \ref{txt:semantics}.
In rule 1, the action $\bm{\phi}[t\leadsto\textit{Id}]$ saves the current package name $\textit{Id}$.
In rule 2, $\bm{\textbf{C}}\cup\{\langle\bm{\phi}[t]\textbf{.}\bm{\theta}[t]\rangle\}$ means concatenating current package name and class name with a dot "$.$" and adds it into the set of classes.
Similarly, rule 3 and 4 extract the member fields and methods in the current class.
Rules 5 and 6 specify the semantic actions for generic types:
creating the mapping $\langle\mathit{Param}\mapsto\mathit{Type}\rangle$ into $\bm{\Omega}$ for bounded generic types (rule 5.1), or
converting the class to \texttt{Object} for unbounded generic types (rule 5.2).
When encountering type parameters, the parser converts them to their bound type names (rule 6).

\subsection{The CSV List}

\begin{figure}[h]
\begin{lstlisting}[language=csv]
Landroid/content/Context;->MODE_WORLD_READABLE:I,public-api,sdk
Landroid/content/Context;->canStartActivityForResult()Z,max-target-r
Landroid/content/Context;->destroy()V,blocked
Landroid/content/Context;->getString(I[Ljava/lang/Object;)Ljava/lang/String;,public-api,sdk
Landroid/content/Context;->getSystemService(Ljava/lang/Class;)Ljava/lang/Object;,public-api,sdk
\end{lstlisting}
    \caption{Sample lines from the \texttt{hiddenapi-flags.csv} file}
    \label{fig:hiddenapi-flags-csv}
\end{figure}

Starting from Android 9, a series of APIs, called \textit{non-SDK interfaces}, were restricted in usage from normal Android apps \cite{web:android/non-sdk-api}.
They are internally used by the Android system and are subject to change without notification.
Android provides a static checker \texttt{veridex} \cite{web:android/veridex} for scanning non-SDK interface usages in Android apps.
It requires a CSV file \texttt{hiddenapi-flags.csv}, in which APIs are listed along with their blocking policies (e.g., public, conditional blocking).

Figure \ref{fig:hiddenapi-flags-csv} displays an excerpt from the CSV file.
Each line corresponds to a specific method or field, i.e., an API.
The tags associated with each API specify its blocking policy.
For examples, the first line indicates that the field \texttt{MODE\_WORLD\_READABLE} is public.
The second line suggests that the API method \texttt{canStartActivityForResult()} can only be used up to Android 11 (code name \underline{r}ed velvet cake).
The third line signifies that API \texttt{Context.destroy()} cannot be invoked in any versions.
It is important to note that \textbf{this file only contain bytecode-level signatures}, which differ from the TXT list presented in Figure \ref{fig:current-txt}.

\subsubsection{API Extraction}
In the CSV file, each line is a JNI signature of either a method or a field.
We extracted their declaring classes from the signatures to obtain the API classes in the CSV list.
Also note that, in Android 9 (API-level 28), the non-SDK interfaces were declared in multiple non-CSV files.
We simply merge all the files to obtain the CSV list for this Android version.

A major difference between the CSV list and other AALs is that the former contains all compiled classes during building Android SDK.
Thus, the CSV list involves classes or methods that were synthesized from high-level Java language features.
For example, anonymous class and lambda expression are two frontend Java expressions that help developers write concise source code.
During compilation, they will be synthesized into some classes or methods.
These synthesized APIs are implicitly called in Java bytecode, which will never be used by Android apps.
To filter out these APIs, we adopt some heuristic rules.
For example, synthesized APIs typically follow some naming patterns (e.g., postfix \texttt{Lambda\textbackslash d+}), which can be used to quickly identify them.
We also excluded the synthesized class initialization methods \cite{web:java/jvm2-2.9}, which have a special method name \texttt{<clinit>}.
On average, these synthesized APIs account for 8.7\% of APIs in the CSV lists.



\section{Empirical Study}
\label{sec:empirical-study}

\subsection{Research Questions}

\begin{figure}[t]
    \centering
    \includegraphics[width=\columnwidth]{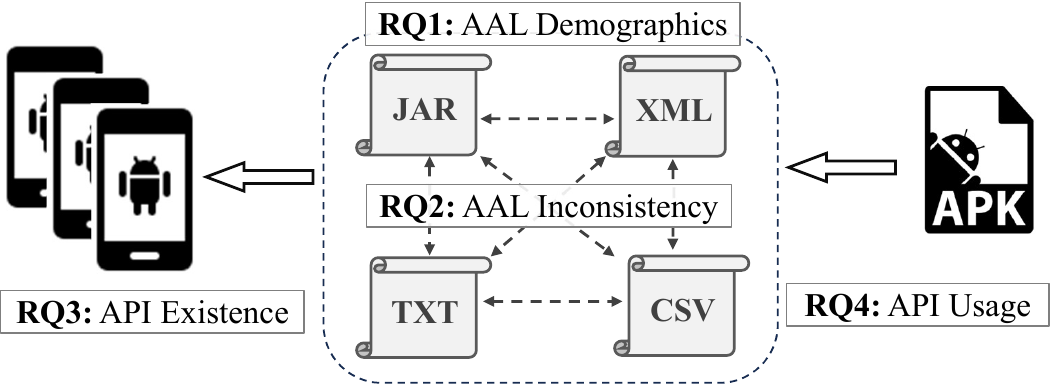}
    \caption{Research questions of our empirical study}
    \label{fig:rq-relationship}
\end{figure}

This study focuses on the completeness and correctness of the four AALs.
To gain a deeper understanding, we also study the API existence on real Android devices and API usage in real-world Android apps.
To drive our study, we raise the following research questions (Figure \ref{fig:rq-relationship}):

\begin{itemize}
    \item \textbf{RQ1 (AAL Demographics):} How many APIs are there in the four AALs? How do they change throughout the evolution of the Android platform?
    \item \textbf{RQ2 (AAL Inconsistency):} What are the differences between APIs included in the four AALs?
    \item \textbf{RQ3 (API Existence):} Do the APIs in AALs really exist on real Android devices?
    \item \textbf{RQ4 (API Usage):} Which APIs in the AALs are commonly used by real-world Android apps?
\end{itemize}

In the subsequent parts, we will provide a detailed explanation of our approach to investigate each RQ and present a thorough discussion of the findings obtained throughout this process.

\subsection{AAL Files Collection}
\label{ssec:preprocessing}

Our study relies on authenticated ALLs.
To ensure this goal, we directly compiled them from Android source code at the release branches of each Android versions.
Specifically, the JAR and XML lists are built from the \texttt{sdk} target and the CSV lists are made with the corresponding build command \cite{web:android/non-sdk-api}.
The TXT lists can be obtained directly from the specific file directory in the codebase \cite{web:android/current-txt}.
As non-SDK interfaces restriction was starting from API-level 28, we only collected the four AALs since that.
To ensure consistency, we limit our analysis to Android 13 (API-level 33), which is the highest API level supported by our available physical devices (Section \ref{sssec:rq3-devices}).

\subsection{RQ1: AAL Demographics}

\begin{table*}[t]
    \centering
    \caption{Numbers of Java API in each AAL}
    \label{tab:api-numbers}
    \resizebox{0.96\textwidth}{!}{
\begin{threeparttable}

\begin{tabular}{@{}cccccccccccccc@{}}
\toprule
\textbf{}            & \textbf{}          & \multicolumn{4}{c}{\textbf{\#Class}} & \multicolumn{4}{c}{\textbf{\#Field}} & \multicolumn{4}{c}{\textbf{\#Method}} \\
\cmidrule(lr){3-6} 
\cmidrule(lr){7-10} 
\cmidrule(l){11-14} 
\textbf{Branch Name} & \textbf{API} & JAR      & XML      & TXT      & CSV      & JAR      & XML      & TXT      & CSV      & JAR       & XML      & TXT      & CSV      \\
\midrule

\texttt{pie-release} & 28 & 4,332 & 4,332 & 4,330 & 10,465 & 22,395 & 22,395 & 22,389 & 70,742 & 44,014 & 35,144 & 38,772 & 63,361 \\
\texttt{android10-release} & 29 & 4,530 & 4,526 & 4,526 & 20,013 & 23,396 & 23,396 & 23,396 & 144,518\upgreenarrow & 46,242 & 36,931 & 40,389 & 205,333\upgreenarrow \\
\texttt{android11-release} & 30 & 4,746 & 4,831 & 4,742 & 22,407 & 25,144 & 25,175 & 25,144 & 155,610 & 48,101 & 38,888 & 41,988 & 228,343 \\
\texttt{android12-release} & 31 & 5,069 & 5,127 & 3,133\downredarrow & 26,661 & 26,492 & 26,512 & 20,254\downredarrow & 179,158 & 50,508 & 40,578 & 27,635\downredarrow & 262,447 \\
\texttt{android12L-release} & 32 & 5,075 & 5,133 & 3,139 & 26,847 & 26,545 & 26,565 & 20,307 & 180,796 & 50,550 & 40,603 & 27,662 & 264,074 \\
\texttt{android13-release} & 33 & 5,276 & 5,334 & 3,197 & 28,906 & 27,868 & 27,888 & 20,669 & 193,357 & 52,716 & 42,371 & 28,178 & 281,872 \\

\bottomrule
\end{tabular}

\begin{tablenotes}
    \item ~~~~~The {\downredarrow} symbol indicates the number of API decreases compared to the previous API-level, while {\upgreenarrow}  means it increases by more than twice.
\end{tablenotes}

\end{threeparttable}
}
\end{table*}

To answer RQ1, we directly counted the numbers of APIs after extraction and preprocessing.
Then, we inspected the results and analyzed how each AAL changes along with Android evolution.
The data is present in Table \ref{tab:api-numbers}.
Each row shows the name of the release branch for the given API-level, and the numbers of three types of APIs in the four AALs, respectively.

Table \ref{tab:api-numbers} shows that the numbers of APIs are mostly increasing during the evolution of the Android platform.
For the CSV list, at API-level 29, the number of three APIs increased by more than double.
This is because, at API-level 28, the CSV list only contained blocked APIs under three restriction levels.
However, starting from API-level 29, this list was expanded to include all bytecode generated from the entire Android framework codebase, including public APIs and whitelist system APIs, etc.
As a result, the APIs contained in the list have significantly grown.
Note that the CSV list contain a considerably larger number of APIs than the others.
We will further investigate its reason in RQ2.

\begin{table}[t]
    \centering
    \caption{Changes of API numbers in TXT lists}
    \label{tab:txt-change}
    \resizebox{0.8\columnwidth}{!}{
\begin{tabular}{@{}ccccccc@{}}
\toprule
      & \multicolumn{2}{c}{\bf \#Class} & \multicolumn{2}{c}{\bf \#Field} & \multicolumn{2}{c}{\bf \#Method} \\
\cmidrule(lr){2-3} \cmidrule(lr){4-5} \cmidrule(l){6-7} 
      & $\boldsymbol{+}$ & $\boldsymbol{-}$ & $\boldsymbol{+}$ & $\boldsymbol{-}$ & $\boldsymbol{+}$ & $\boldsymbol{-}$ \\
\midrule

~~~~28~\ding{213}~29 & 196 & 0 & 1,028 & 21 & 2,009 & 392 \\
~~~~29~\ding{213}~30 & 216 & 0 & 1,756 & 8 & 1,599 & 0 \\
~~~~30~\ding{213}~31 & 198 & 1,807 & 1,150 & 6,040 & 1,335 & 15,688 \\
~~~~31~\ding{213}~32 & 6 & 0 & 53 & 0 & 27 & 0 \\
~~~~32~\ding{213}~33 & 128 & 70 & 980 & 618 & 1,085 & 569 \\

\bottomrule
\end{tabular}
}
\end{table}

However, for the TXT list at API-level 31, the number of classes decreases by 33.9\% ($=(4742-3133)/4742$), along with a significant drop in the numbers of fields and methods.
Meanwhile, the other AALs still increased their API numbers.
This observation suggests that the reduction in API numbers within the TXT list may not attribute to the removal of APIs from the Android framework.
Table \ref{tab:txt-change} shows the numbers of added/removed APIs (by pairwise set difference) in the TXT list.
At API-level 31, there were 1,807 classes, 6,040 fields and 15,688 methods being removed from the TXT list.
Moreover, 6,038 (99.97\%) removed fields and 15,682 (99.96\%) moved methods come from these removed classes.
According to their package names, we aggregated the 1,807 classes and found that 78.1\% of them come from either JDK (under package \texttt{java.*} or \texttt{javax.*}) or third-party packages (in \texttt{org.*}).
This means, at API-level 31, most non-Android packages were removed from the TXT list, causing a significant drop in the numbers of its including APIs.

\begin{tcolorbox}
\textit{\textbf{\newfinding}}
For AALs under study, the rules governing API inclusion can undergo changes throughout the evolution of the Android platform.
Consequently, the number of APIs in AALs may experience significant increases or decreases.
\end{tcolorbox}

For backward-compatibility purpose, Android API removal is carried out carefully~\cite{cda-msr18}.
This is also supported by Table \ref{tab:txt-change}, such that added APIs are generally more than removed APIs.
However, it turns out that APIs removed from AALs can still exist. 
Taking the 569 removed methods from 32 to 33 in TXT (the last cell in Table \ref{tab:txt-change}) as an example.
For each method, we search the Android codebase for API-level 32 and 33, and compare its presenting status in both versions.
Among them, 547 (96.1\%) methods were actually removed from the codebase.
For the remaining 22 methods, four of them were labeled by a \texttt{@removed} annotation and became non-SDK interfaces, while the other 18 did not exist in either API-level 32 or 33.
An example of the former case is \texttt{WebSettings.setAppCacheMaxSize(long)}, which was labeled as deprecated in 32 and later annotated with \texttt{@removed} in its Javadoc string in 33.
We successfully invoke this method on a Pixel 3 XL device running Android 13 via Java reflection.
This indicates that APIs removed from AALs can still be accessed by apps, posing a soundness challenge to API-based compatibility issue detectors.


\begin{tcolorbox}
\textit{\textbf{\newfinding}}
APIs removed from the AALs may still exist in the Android source code or runtime. 
\end{tcolorbox}

The findings above imply that there can be a disparity between the APIs declared in AALs and those accessible to developers.
To gain a deeper understanding of this situation, we explored the differences among the four AALs (RQ2) and assess their presence in real devices (RQ3).
Finally, we estimate to what extent these AALs cover API usages in real-world apps (RQ4).

\subsection{RQ2: AAL Inconsistency}

Table \ref{tab:api-numbers} demonstrates that, even at the same API-level, different AALs exhibit great differences in the numbers of their including APIs.
For instance, at API-level 33, the number of classes in the CSV list is nine times greater than that in the TXT list.
Additionally, the other three AALs contain significantly fewer APIs compared to the CSV list.
RQ2 aims to analyze the differences among AALs in a qualitative and quantitative way.

\begin{figure}[t]
    \centering
    \includegraphics[width=\columnwidth]{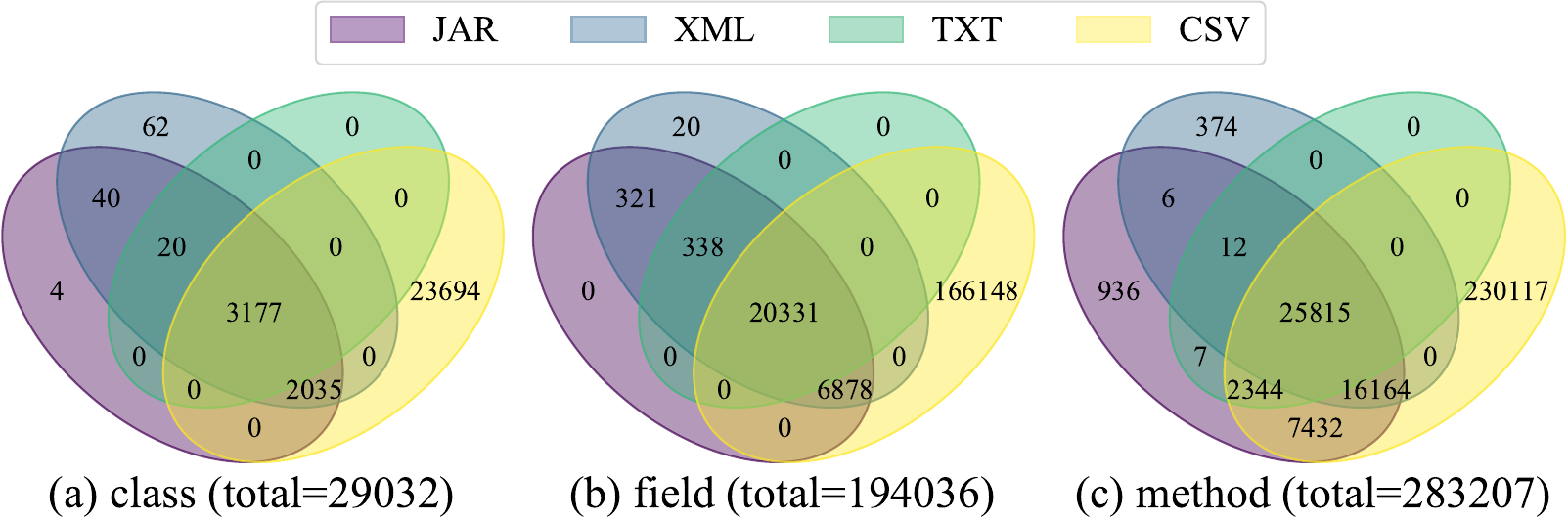}
    \caption{API differences between each AALs (API-level 33)}
    \label{fig:api-differences}
\end{figure}

Figure \ref{fig:api-differences} visualizes the differences of four AALs at API-level 33.
For example, Figure \ref{fig:api-differences}a shows that, among the 29,032 classes declared in the four AALs, only 3,177 (10.9\%) classes are included by all of them.
As for fields and methods, the percentages are 10.5\% and 9.1\%.
The XML list has 62 classes that do not exist in the other three AALs, 40 classes only appear in it and the JAR list, and so on.
One can easily verify the data by summing up all numbers in an ellipse to obtain the number of the corresponding API in the last row of Table \ref{tab:api-numbers} (e.g., for CSV classes, 3,177+2,035+23,694=28,906).

There are only a few APIs in common among the four AALs.
The 20,331 shared fields and 25,815 shared methods all come from the 3,177 shared classes.
Among these classes, 3,153 (99.2\%) of them are located in the \texttt{android.*} namespace.
These classes provide fundamental system functionalities to the Android apps, like Android app components, UI widgets, telephony, network management, and so forth.
It indicates these shared APIs are the core Android APIs.
The other 13 classes were repackaged from Apache HttpClient 4.0 since Android 1.0 \cite{web:apache/android-httpclient} and are under the \texttt{org.apache.http.*} package.
The remaining 11 classes are Khronos OpenGL interfaces \cite{web:khronos/egl} that are located in the \texttt{javax.microedition.khronos.*} package.

\begin{tcolorbox}
\textit{\textbf{\newfinding}}
At API-level 33, only about 10\% of all Java APIs are shared by all the four AALs.
\end{tcolorbox}

Figure \ref{fig:api-differences} shows that the CSV list has a significant number of exclusive APIs.
Particularly, 23,694/29,032 = 81.6\% of classes, 85.6\% of fields, and 81.3 \% of methods are exclusive to this list among the four AALs.
Furthermore, we found that 134,062 (80.7\%) of the exclusive fields and 204,467 (88.9\%) of the exclusive methods belong to these 23,694 classes.
The result suggests that a large proportion of CSV APIs come from the classes that are only present in it.

To comprehend the characteristics of these exclusive classes, due to the lack of explicit categorization rules, we performed manual inspection on them.
Specifically, from the total population of 23,694 classes, a sample of 379 classes was randomly selected to achieve representativeness with a confidence level of 95\% and a margin of error of 5\% \cite{wang2021understanding}.
For each class under inspection, an author searched it on the Android Code Search website\footnote{\url{https://cs.android.com/}} and collected all characteristics about this class (e.g., access level, special annotations).
Then, the author determined the major reason why this class is present only in the CSV list.
For example, a class can be private and deprecated at the same time, and we consider the former to be the major reason since deprecated APIs can also present in the other three AALs.
An author performed this process, and another author verified the result.
In case of conflicts, they searched back to the source code and discussed until a consensus was reached.
The final result is shown in Table \ref{tab:csv-exclusive-sample}.
There are three types of exclusive classes in the sample:

\begin{table}[t]
    \centering
    \caption{Categorization of 379 exclusive classes in CSV list}
    \label{tab:csv-exclusive-sample}
    \resizebox{\columnwidth}{!}{\begin{tabular}{l|c|l|c}
\hline\hline
\multirow{4}{*}{\begin{tabular}[c]{@{}l@{}}Intentionally hidden\\  from public APIs\end{tabular}}       & \multirow{4}{*}{151} & with \texttt{@hide} annotations                                                                                            & 74 \\ \cline{3-4} 
                                                                                                        &                      & with \texttt{@removed} annotations                                                                                          & 2  \\ \cline{3-4} 
                                                                                                        &                      & non-public access levels                   & 17 \\ \cline{3-4}
                                                                                                        &                      & internal Android packages                                                                                                 & 58 \\ \hline\hline
\multirow{4}{*}{\begin{tabular}[c]{@{}l@{}}Compiled from foreign\\  programming languages\end{tabular}} & \multirow{4}{*}{115} & from \texttt{.aidl} files                                                                                                 & 90 \\ \cline{3-4} 
                                                                                                        &                      & from \texttt{.proto} files                                                                                                & 16 \\ \cline{3-4} 
                                                                                                        &                      & from C/C++ header files                                                                                                   & 8  \\ \cline{3-4} 
                                                                                                        &                      & from \texttt{.sysprop} files                                                                                              & 1  \\ \hline\hline
\multirow{2}{*}{\begin{tabular}[c]{@{}l@{}}Repackaged from other\\  Java libraries\end{tabular}}        & \multirow{2}{*}{113} & JDK classes                                                                                                               & 32 \\ \cline{3-4} 
                                                                                                        &                      & third-party libraries                                                                                                                    & 81 \\ \hline\hline
\end{tabular}}
\end{table}

\begin{itemize}
\item 151 classes are explicitly defined as ``inaccessible''.
They can be attached with \texttt{@hide} or \texttt{@removed} annotations, or have the three non-public access levels in Java.
Alternatively, 58 classes are defined in internal packages (e.g., \texttt{com.android.internal.*}).

\item 115 classes are compiled from languages other than Java.
Among them, 90 classes are generated from AIDL interfaces (\texttt{.aidl} files), while 16 classes are from Protocol Buffers message definitions (\texttt{.proto} files).
Besides, 9 classes are transcompiled from C/C++ header files or \texttt{.sysprop} file.

\item 113 classes are repackaged from external Java libraries.
Nearly one-third of them come from JDK (e.g., NIO classes, concurrent collections).
The other classes originate from third-party libraries. 
\end{itemize}

\begin{tcolorbox}
\textit{\textbf{\newfinding}}
The CSV list includes a large number of APIs which do not exist in the other three AALs. They are mostly generated from other programming languages, repackaged from external libraries, or intentionally blocked from Android SDK.
\end{tcolorbox}

\begin{figure}[t]
    \centering
    \includegraphics[width=\columnwidth]{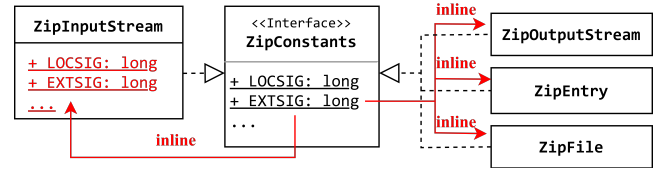}
    \caption{Illustration of fields inlined from \texttt{ZipConstants}}
    \label{fig:zip-inlined}
\end{figure}

The other AALs also declare exclusive APIs.
For example, the XML list contains 20 fields and 374 methods that do not present in the other AALs.
All of them come from the 62 exclusive classes, whose package names all start with \texttt{android.test.*} or \texttt{junit.*}.
However, we searched the classes in other AALs and could not find APIs under such packages.
This indicates the XML list is the only AAL that involves testing-related APIs.
The JAR list has 4 exclusive classes and 936 exclusive methods.
Upon examination, We confirmed that the four classes are all annotation definitions.
Among the exclusive methods, 866 are default constructors generated by the Java compiler, which are absent from the source code.
The remaining 70 methods are inlined from their super classes in the JAR list.
For example, 49 out of the 70 exclusive methods belong to the class \texttt{RecordingCanvas} in JAR, while none of the methods are defined in this class.
Instead, they are defined in its super class \texttt{Canvas}.
Another such case will be illustrated in Figure \ref{fig:zip-inlined}.

\begin{tcolorbox}
\textit{\textbf{\newfinding}}
Testing modules contribute to all the exclusive APIs in  XML,
while exclusive methods in  JAR  are either default constructors or methods inlined from the super classes.
\end{tcolorbox}

Although the CSV list has a large number of APIs, it, however, cannot cover them all.
For example, 40 classes are only present in the JAR and  XML lists; 20 classes are shared by the other three AALs.
A common characteristic of them is that, they are all empty classes.
In other words, these classes have no field or method members, and thus cannot be shown in the CSV list.

As for the fields and methods not covered by CSV, we compared their API signatures and the source code, and found these members were inlined to the concrete classes while their abstract super classes were hidden in the AALs.
Figure \ref{fig:zip-inlined} illustrates such an example.
The class \texttt{ZipInputStream} (and \texttt{ZipOutputStream}, \texttt{ZipEntry}, etc.) implements the interface \texttt{ZipConstants}, which has multiple constant fields.
The CSV list also declares these fields to be members of \texttt{ZipConstants}.
However, in JAR and XML, these fields are inlined to the concrete classes that implement this interface, as indicated by the red arrows.
Consequently, these constant fields become the members of the sub-classes in these two AALs.
Similar to fields, the inlined methods also cannot be covered by CSV.

\begin{tcolorbox}
\textit{\textbf{\newfinding}}
Empty classes cannot be represented in the CSV list.
Some fields and methods are inlined to the concrete classes, which also prevents their inclusion in the CSV list.
\end{tcolorbox}

Due to space limit, this paper only displays the results at API-level 33.
However, we also observed similar findings in the other API-levels.
The complete results can be found in our released dataset.

\subsection{RQ3: API Existence}

\subsubsection{Method}
\label{sssec:rq3-setup}

To study the API existence in real Android devices, we developed an Android app, \textsc{AAL-Reflector}.
It accepts the AALs at the corresponding runtime version and tries to access their APIs using \textit{direct Java reflection} \cite{li2019understanding}.
This means \textsc{AAL-Reflector} only retrieves APIs explicitly defined in a class, without searching for inherited APIs from its parents.
If an AAL API cannot be reflected, \textsc{AAL-Reflector} regards it as a non-existed API.

\textsc{AAL-Reflector} also counts the numbers of \textit{non-AAL} fields and methods.
Here, ``non-AAL fields'' refer to those that are listed through the reflection API \texttt{Class.getDeclaredFields()} but do not present in any of the four AALs.
Similarly, non-AAL methods are also counted.
To find non-AAL APIs, we only use classes in the four AALs, and do not search for ``non-AAL classes''.
This is because Java class loading is an on-demand process, one cannot enumerate all loadable classes without prior knowledge (Java package name, etc.) or access to the system images (it requires rooted devices) \cite{web:stackoverflow/load-android}.

\subsubsection{Devices Pool}
\label{sssec:rq3-devices}

We run \textsc{AAL-Reflector} on multiple Android devices.
Specifically, six physical devices with customized operating systems were available to us, which include three Android versions.
For each available Android version, we also created a virtual device running the corresponding stock Android image.
In total, nine Android devices were used in studying RQ3.

\subsubsection{Result}

\begin{table*}[t]
    \centering
    \caption{Numbers of non-existed APIs on Android devices (retrieved by direct Java reflection)}
    \label{tab:missing-api-numbers}
    \resizebox{0.95\textwidth}{!}
    {
\begin{threeparttable}

\begin{tabular}{@{}ccccccccccccc@{}}
\toprule
           &          &                     & \multicolumn{5}{c}{\textbf{Fields}} & \multicolumn{5}{c}{\textbf{Methods}} \\ \cmidrule(lr){4-8} \cmidrule(l){9-13} 
\textbf{Device}     & \textbf{System}   & \textbf{API} & JAR   & XML   & TXT  & CSV  & \textit{non-AAL}$^\dag$  & JAR   & XML   & TXT   & CSV  & \textit{non-AAL}$^\dag$  \\ \midrule

Redmi Note 10       & MIUI 12.0.3       & 30 & 655 & 673 & 655 & 106    & 559/1,116 & 767 & 375 & 31 & 279 & 1,007/1,419 \\
Samsung Galaxy A50s & One UI 3.1        & 30 & 655 & 673 & 655 & 108    & 7,622/12,657 & 767 & 375 & 31 & 311 & 8,519/10,525 \\
virtual             & stock Android     & 30 & 655 & 673 & 655 & 85     & 81/195 & 767 & 375 & 31 & 209 & 189/281 \\
\midrule
Redmi 10X           & MIUI 13.0.1       & 31 & 655 & 675 & 334 & 17     & 470/878 & 903 & 386 & 13 & 59 & 1,003/1,227 \\
vivo Y33s           & OriginOS ocean    & 31 & 655 & 675 & 334 & 58     & 1,283/2,951 & 903 & 386 & 13 & 72 & 3,392/3,931 \\
virtual             & stock Android     & 31 & 655 & 675 & 334 & 4      & 7/10 & 903 & 386 & 13 & 3 & 9/14 \\
\midrule
Redmi Note 12       & MIUI 14.0.1       & 33 & 659 & 679 & 338 & 57     & 634/1,378 & 961 & 392 & 19 & 1,776 & 2,869/3,184 \\
Samsung Galaxy A53  & One UI 5.1        & 33 & 659 & 679 & 338 & 255    & 7,613/12,458 & 961 & 392 & 19 & 813 & 8,756/11,009 \\
virtual             & stock Android     & 33 & 659 & 679 & 338 & 235    & 406/572 & 961 & 392 & 19 & 784 & 537/669 \\

\bottomrule
\end{tabular}

\begin{tablenotes}
    \item $^\dag$``non-AAL'' are APIs in the devices but do not present in all AALs. Item $x/y$ means there are $x$ public APIs and totally $y$ APIs.
\end{tablenotes}

\end{threeparttable}
}
\end{table*}

Table \ref{tab:missing-api-numbers} presents the list of our devices and the results of running \textsc{AAL-Reflector} on them.
The numbers under each AAL are the number of APIs included in the AAL but failed to reflect on the device.
Additionally, the column ``non-AAL'' shows the numbers of non-AAL fields or methods (Section \ref{sssec:rq3-setup}), in terms of public APIs and all APIs, respectively.

Table \ref{tab:missing-api-numbers} shows that the API existing statuses for JAR, XML, and TXT are consistent across Android devices at the same API-level.
For example, all Android 11 devices lack 655 fields and 767 methods declared in the JAR list, no matter whether they were customized systems or stock Android.
Interestingly, \textbf{their non-existed APIs are those not covered by the CSV list}.
Take the JAR fields at API-level 33 as an example.
Table \ref{tab:missing-api-numbers} tells that there are 659 fields declared in the JAR list but cannot be retrieved on all Android 13 devices.
Notably, $659=321+338$, which is exactly the JAR fields that have not been included by the CSV list (Figure \ref{fig:api-differences}b).
The same situation also applies to the non-existed APIs in XML and TXT.
This is not surprising.
We already pointed out that, these APIs either come from the test framework (Finding 5) or were inlined to their concrete classes in the AALs (Finding 6).
They do not necessarily represent the actual APIs in Android devices.

On the other hand, only a few CSV APIs are absent on the devices.
For instance, there are 281K CSV methods at API-level 33 (Table \ref{tab:api-numbers}), of which only 784 (0.28\%) do not exist in the stock Android 13 system.
Customized systems lack more CSV methods than stock Android.
Moreover, there is no significant overlap between the non-existed CSV APIs on customized systems.
The two customized Android 13 devices have 57 and 255 CSV fields, respectively, that are not in their system images.
However, only 7 of them are in common. Similarly, only 41 non-existed CSV methods are shared by both devices.
Since most CSV APIs are declared as \textit{non-SDK interfaces}, it indicates that customized systems can modify these restricted APIs while keeping the core Android APIs consistent.

\begin{tcolorbox}
\textit{\textbf{\newfinding}}
The CSV APIs present discrepancies for their existences in different Android systems.
\end{tcolorbox}

Non-AAL APIs exist on the real devices, but cannot be found in any of the four AALs.
Customized devices contain more such APIs than stock Android devices.
Moreover, Table \ref{tab:missing-api-numbers} shows that most of them have a \textit{public} access level.
However, there is no public documentation on the open web for these non-AAL APIs.
In fact, most of these APIs are used by device manufacturer apps on the corresponding systems \cite{web:github/samsung-customized-keyevent}.
For example, there is a non-AAL method \texttt{Looper.isPerfLogEnable()} in the Samsung Galaxy A53.
We found it raises a null-pointer exception when the user encounters a connectivity issue \cite{web:samsung/getperflogenable}.

\begin{tcolorbox}
\textit{\textbf{\newfinding}}
Customized Android devices widely define their own Java APIs in the Android framework classes.
Most of them have a public access level.
\end{tcolorbox}

\subsection{RQ4: API Usage}

\subsubsection{Method}

We implemented a tool, \textsc{APK-Analyzer}, to analyze API usages in an app, i.e., an APK file.
The tool integrates Soot \cite{lam2011soot} and veridex \cite{web:android/veridex}, and performs an intra-procedural flow-insensitive analysis.
In particular, it scans a method body to collect three types of API callsites in the application classes:
\begin{enumerate}
    \item \textit{Direct API call}: referencing APIs via \texttt{FieldRef} or \texttt{InvokeExpr} Jimple expressions;
    \item \textit{Extra API call}: using APIs under Android packages but not present in any of the four AALs;
    \item \textit{Reflect API call}: calling APIs via Java reflection. Here we directly parse the output of veridex and filter out our desired API usages.
\end{enumerate}

The analysis of types (2) and (3) allows us to recover more used APIs beyond those listed in the AALs.
Also note that, as JDK classes (e.g., generic collections) can be included by AALs, \textsc{APK-Analyzer} will discard them from the output as they are out of our interests.

\subsubsection{App Dataset}

We collected APK files from three data sources, each of them representing a typical type of Android app. They are:

\begin{itemize}
    \item \textbf{Open-source app}. F-Droid is an open-source app catalog. It has been widely used in Android-related studies \cite{aper-icse2022}. We downloaded all the 4,090 available apps from F-Droid.
    \item \textbf{Commercial app}. Google Play consists of millions of commercial Android apps \cite{cao2022rotten}. To ensure representativeness, we only collected the top-500 apps in 32 categories indexed by AppBrain\footnote{\url{https://www.appbrain.com/apps/popular}}. Finally, we downloaded 15,334 Google Play apps.
    \item \textbf{Android malware}. Malware can exhibit distinguishable API usages to normal apps \cite{tam2017evolution}. Here we directly used the sample of 1,238 malicious apps collected by Cao et al. \cite{cao2022rotten}.
\end{itemize}

For the collected apps, we discarded those that have not invoked Android APIs or cannot be analyzed by \textsc{AAL-Analyzer} (e.g., obfuscated APKs \cite{martin2016survey}).
After app exclusion, 4,046 open-source apps, 12,968 commercial apps, and 745 malicious apps remain.
In total, 17,759 real Android apps built up our experimental dataset.

\subsubsection{Result}

\begin{table}[t]
    \centering
    \caption{Android API usages in the app datasets}
    \label{tab:call-api-total}
    \resizebox{\columnwidth}{!}{\begin{threeparttable}

\begin{tabular}{@{}cccccccc@{}}
\toprule
            &       & \multicolumn{2}{c}{\textbf{direct call}} & \multicolumn{2}{c}{\textbf{extra call}} & \multicolumn{2}{c}{\textbf{reflect call}} \\
\cmidrule(lr){3-4} \cmidrule(lr){5-6} \cmidrule(l){7-8} 
\textbf{Dataset}     & \#APK & \#API       & \#APK      & \#API      & \#APK      & \#API       & \#APK       \\
\midrule

F-Droid & 4,046 & 24,973 & 4,046 & 128,233 & 3,332 & 487 & 1,601 \\
Google Play & 12,968 & 28,200 & 12,968 & 712,932 & 12,719 & 1,556 & 11,609 \\
Malware & 745 & 11,854 & 745 & 12,129 & 629 & 230 & 360 \\

\bottomrule
\end{tabular}

\begin{tablenotes}
    \item ``\#API'' stands for all the used APIs (i.e., fields plus methods); ``\#APK'' is the number of APKs that use certain calls of API.
\end{tablenotes}

\end{threeparttable}
}
\end{table}

Table \ref{tab:call-api-total} shows the overall API usages, categorized into three types of API calls.
For example, among the 4,046 F-Droid apps, all of them directly invoke APIs in the four AALs, with a total number of 24,973 used APIs (fields and methods).
Only 1,601 of them use reflection to invoke a total number of 487 APIs, etc.

All apps involve direct API calls, which means every app invokes at least one AAL-included API (either field or method).
Further, we found that most of the directly-called APIs are those shared by the four AALs.
Take F-Droid as an example, in which 24,973 AAL APIs (3,291 fields and 21,682 methods) are used.
Among them, 1,738 (52.8\%) fields and 15,825 (73.0\%) methods appear in all the four AALs.
Meanwhile, 1,537 (46.7\%) fields and 3,926 (18.1\%) methods are only present in CSV.
Apart from them, F-Droid apps also invoke 6 JAR-only and 27 XML-only methods.
Notably, the inlined members in JAR (Finding 5) are used by real-world apps.
Accessing inlined members from bytecode is valid because their access can be dynamically dispatched at runtime.
However, it also poses a challenge for static analyzers to resolve the actual APIs in the relying AALs if they do not involve such inlined members (i.e., CSV).
The results on Google Play apps are similar to F-Droid.
As for malware, almost 90\% of the direct API calls are for shared APIs.

\begin{tcolorbox}
\textit{\textbf{\newfinding}}
Most of the AAL API usages are for those common APIs shared among the four AALs.
We also observed non-negligible usages of APIs that are not shared by all the AALs.
\end{tcolorbox}

The number of extra-call APIs is far more than the APIs in AALs.
Most of these APIs are Android Support APIs (package \texttt{android.support.*}) or AndroidX APIs (package \texttt{androidx.*}) \cite{web:android/androidx}, which were packed in APK files to ensure backward API compatibility.
They can be obfuscated by identifier renaming \cite{dong2018understanding}, leading to an explosion of unique API names.
In F-Doird, 97.8\% of the extra call APIs (including obfuscated ones) are either Android Support APIs or AndroidX APIs.
As for Google Play and Malware, the percentages are 92.5\% and 89.8\%, respectively.

Interestingly, we observed the usage of vendor-customized APIs.
To be specific, we dumped the extra call APIs and checked if they had intersections with the non-AAL APIs that were discovered on our experimental devices in RQ3 (Table \ref{tab:missing-api-numbers}).
As a result, we found callsites of three non-AAL APIs in F-Droid apps, and 78 callsites in Google Play apps.
For example, an open-source app, App Manager \cite{web:github/app-manager}, utilizes the field \texttt{MIUI\_OP\_START} to check whether a file operation can be performed on the MIUI system.

\begin{tcolorbox}
\textit{\textbf{\newfinding}}
Extra API calls are mostly for Support and AndroidX APIs.
Apps from F-Droid and Google Play also contain usages of vendor-customized APIs.
\end{tcolorbox}

In our collected apps, 1,601 (39.6\%) F-Droid apps contain reflection calls.
This percentage is less than that in Google Play apps (89.5\%) and Malware (48.3\%).
By comparing to the CSV file, we found that only 12 reflection call APIs in F-Droid are public,
 while in Google Play apps and Malware, the numbers are 35 and 11, respectively.
This implies that most reflection API calls in real-world apps are targeting non-SDK interfaces (i.e., non-public APIs).

\begin{tcolorbox}
\textit{\textbf{\newfinding}}
Commercial apps and malware are more likely to use reflection calls to non-SDK interfaces than open-source apps.
\end{tcolorbox}

\section{Discussion}
\label{sec:discussion}

\subsection{Implications of Our Findings}

\noindent$\bullet$~\textbf{Extracting API knowledge from web documentation or source code can be hard.}
The web documentation is well-known to be unreliable \cite{a3-tse2020,bavota2014impact}.
As for the Android source code, researchers often focus on Java files, while Android APIs can also be generated from other programming languages (Finding 4).
Moreover, it should consider the inlined members, as they are present in the app's bytecode (Finding 9).
Extracting such APIs poses a significant challenge of emulating a compiler.
Android Support and AndroidX libraries should also be taken into account (Finding 10).

\noindent$\bullet$~\textbf{AALs are incomplete, inconsistent, and even unstable.}
None of the AALs can fully cover the other three (Figure \ref{fig:api-differences}).
Even the largest AAL, CSV, lacks testing APIs (Finding 5).
Moreover, the API inclusion policy may change over time (Finding 1).
Therefore, it is important for tools that rely on specific AALs to be mindful of potential API mismatches.
For instance, four methods were removed from the TXT list in Android 13, but they remain accessible from the Android runtime (Finding 2).
Remarkably, we found that 5,055 (39\%) Google Play apps in our dataset invoked these methods.

\noindent$\bullet$~\textbf{APIs in AALs may not exactly mirror the real Android APIs.}
Discrepancy exists between the AAL APIs and those present in Android source code or runtime (Finding 2).
In particular, JAR, XML, and TXT lists contain inlined members (Figure \ref{fig:zip-inlined}), which are not included by CSV.
These inlined APIs are also absent from the Android runtime (Finding 7).
As a result, it is necessary to employ distinct API handling for static and dynamic issue detectors.

\subsection{Suggestions for Future Research}

\noindent$\bullet$~\textbf{Being specific when parsing Android source code.}
The Android codebase is comprised of over one thousand sub-modules and is developed in tens of programming languages.
Thus, parsing the whole codebase to extract Android APIs is not a trivial task.
For reproduction purpose, it would be better if more details about such process is presented.
For examples, which programming languages or Android system sub-modules are considered, or what kinds of pre-processing are conducted (generic type erasing, member inlining, etc.).

\noindent$\bullet$~\textbf{Choosing appropriate AALs based on the tasks.}
The four AALs we studied vary in their API inclusion policies.
Consequently, researchers are encouraged to provide clearer explanations regarding the particular application of AAL within the context of their research.
For example, one may choose the TXT list when they only focus on the core Android functionalities,
use XML if Android testing APIs are needed, or resort to CSV for native/IPC API analysis.

\noindent$\bullet$~\textbf{Seeking further research on non-SDK APIs and/or customized APIs.}
Apart from normal Android APIs, we have also observed usages of non-SDK interfaces and vendor-customized APIs.
They can bring security vulnerabilities \cite{el2021dissecting} and compatibility issues \cite{li-icsme16}, which would ultimately have a negative impact on the end-users.
To enhance the security and privacy of the Android system, more research efforts should be taken to understand these APIs.

\subsection{Threats to Validity}




\noindent$\bullet$~\textbf{Limited coverage of Android versions and devices.}
Our study covers six Android versions and nine physical devices, which may limit the generalizability of our findings.
Although these versions span multiple major Android releases and the devices include both stock Android and vendor-customized systems, they may not fully represent the diversity of the Android ecosystem, especially given the rapid evolution of Android versions and the continuous introduction of new device models.
As a result, some API behaviors or vendor-specific customizations may not be captured in our analysis.

\noindent$\bullet$~\textbf{Restricted accessibility of system-only APIs.}
Certain APIs are only accessible to system apps and cannot be invoked by third-party apps, even via reflection.
Such APIs are therefore considered non-existent by \textsc{AAL-Reflector}, which focuses on APIs accessible to normal applications.
In our analysis, we identified only a small number of such APIs, primarily in the CSV AALs.
Since these APIs are not intended for use by third-party apps and fall outside the scope of most Android app analyses, we did not further investigate them.
Nevertheless, this limitation may affect studies that explicitly target system-level or privileged applications.

\noindent$\bullet$~\textbf{Potential inaccuracies in API usage extraction.}
A potential construct threat arises from the API usage extraction process implemented by \textsc{APK-Analyzer}, which may not perfectly capture all API invocations in an app.
To reduce this threat, we restricted our analysis to application-level classes and relied on conservative extraction rules.
While this approach improves precision, it may underestimate the overall API usage in some apps.
Future work could incorporate more advanced static or hybrid analysis techniques to further improve extraction accuracy.

\section{Related Work}
\label{sec:related-work}

\subsection{Empirical Study about Android APIs}

Android Operating System Project (AOSP) is a large and complicated open-source project \cite{mahmoudi2018android}.
For this reason, many researchers aim to mine empirical knowledge from the Android codebase.
Among them, the very first work was conducted by McDonnell et al. \cite{mcdonnell-icsm2013}.
They studied the SDK API lists (i.e., the XML list) in 17 Android releases and analyzed 10 open-source apps for how they react to API changes.
They have shown that, upon Android 4.2 (API-level 17), Android APIs experienced a fast-evolving process.
They also found that developers tend to avoid frequently upgrading outdated APIs.
Later, Xia et al. \cite{rapid-icse2020} investigated how developers handle incompatible evolving APIs on a dataset of 300,000 apps.
To determine whether developers provide alternative implementations to incompatible APIs, they proposed an SVM-based approach, RAPID.
They found that developers do not always provide alternative implementations to incompatible APIs.
Meanwhile, they are more willing to do so if Android officially documents API replacement recommendations.

Researchers have also studied other categories of Android APIs.
Early in 2016, Li et al. \cite{li-icsme16} studied inaccessible APIs in the Android framework, and found that they were pervasively used. 
These inaccessible APIs were later announced to be part of the non-SDK interfaces and enforced by restrictions \cite{web:android/non-sdk-api}, which form the restricted APIs in the CSV list in our study.
However, Yang et al. \cite{yang-icse2022} suggested that restricted APIs can still be accessed by normal apps.
%
%
%
Besides, Li et al. proposed CDA \cite{cda-msr18} to collect deprecated APIs from Android source code.
Liu et al. \cite{liu-icseseip21} studied silently-evolved methods, which will change their implementation but keep the original documentation.


\subsection{Android Compatibility Issue}

Android compatibility issue is a long-standing problem \cite{cai-issta19}.
%
The most recent tool is \textsc{PSDroid} \cite{psdroid-icse2023}.
It employs a path-sensitive analysis of API invocation with four common API usage check patterns, and harvests incompatible API lists by parsing the TXT list files.
\textsc{SAINTDroid} \cite{saintdroid-dsn22} is another recent tool that aims at detecting three types of API compatibility issues.
It directly parses the Android source code to obtain an API database.
Prior to them, Liu et al. \cite{liu2022automatically} conducted a replicability study about seven Android compatibility issue detectors:
ACID \cite{acid-saner21}, ACRYL \cite{acryl-msr19,acryl-emse2020}, PIVOT \cite{pivot-icse19},  \textsc{CiD} \cite{cid-issta18}, \textsc{IctApiFinder} \cite{ictapifinder-ase18}, \textsc{FicFinder} \cite{ficfinder-ase16,ficfinder-tse2018}, and CIDER \cite{cider-ase18}.
Their study shows that tools with systematic API collecting methods can have more detection results than those without.
This finding highlights the importance of a complete list of incompatible Android APIs.
Our study systematically analyzes the four official AALs, which are complementary to their work.

Compatibility issues can be exposed dynamically.
\textsc{JUnitTestGen} \cite{junittestgen-ase22} is capable of generating compatibility issue-triggering test cases for Android.
It accepts an arbitrary API list and synthesizes API usages as unit test cases.
Researchers have also proposed methods to automatically repair such API-induced compatibility issues: \textsc{RepairDroid} \cite{repairdroid-icse22} uses a domain-specific language for describing repairing strategies, \textsc{AppEvolve} \cite{appevolve-issta19} updates API usages from example patches.

Besides API incompatibility, Android compatibility issues can manifest themselves in other aspects, such as UI inconsistencies across running devices \cite{mimic-icse19},  different handling methods for configuration items \cite{confdroid-ase21,conffix-issta23}, changes in the system's permission mechanism \cite{aper-icse2022}, etc.

\section{Conclusion}
\label{sec:conclusion}

This work starts from a systematic literature review about Android API research, from which we summarized four kinds of commonly-used official Android API Lists (AALs).
In a comprehensive empirical study of the four AALs, we investigated the API characteristics, differences, existence on real devices, and usage in real-world apps.
Based on our empirical findings, we have drawn several implications for future work in related areas.
Our study reveals the instability and inconsistency between the four AALs.
It also points out research opportunities about non-SDK APIs and vendor-customized APIs in the Android ecosystem.
Our study serves as valuable resources for practitioners and researchers, offering them a deeper understanding of the Android APIs and the API lists.


\begin{acks}
The authors sincerely appreciate all anonymous reviewers for their valuable comments, which have helped improve this paper.
This work is supported by the National Natural Science Foundation of China (Grant Nos. 62372219 and 62202213).
\end{acks}

\bibliographystyle{ACM-Reference-Format}
\bibliography{pub,web}

@inproceedings{acid-saner21,
  title={Android compatibility issue detection using api differences},
  author={Mahmud, Tarek and Che, Meiru and Yang, Guowei},
  booktitle={2021 IEEE International Conference on Software Analysis, Evolution and Reengineering (SANER)},
  pages={480--490},
  year={2021},
  organization={IEEE}
}

@inproceedings{acryl-msr19,
  title={Data-driven solutions to detect api compatibility issues in android: an empirical study},
  author={Scalabrino, Simone and Bavota, Gabriele and Linares-V{\'a}squez, Mario and Lanza, Michele and Oliveto, Rocco},
  booktitle={2019 IEEE/ACM 16th International Conference on Mining Software Repositories (MSR)},
  pages={288--298},
  year={2019},
  organization={IEEE}
}

@inproceedings{li2023towards,
  title={Towards the Adoption and Adaptation of the AndroidX Library: An Empirical Study},
  author={Li, Jiacheng and Huang, Kerui and Wang, Sinan and Liu, Yepang},
  booktitle={2023 IEEE 23rd International Conference on Software Quality, Reliability, and Security (QRS)},
  pages={418--427},
  year={2023},
  organization={IEEE}
}

@article{acryl-emse2020,
  title={API compatibility issues in Android: Causes and effectiveness of data-driven detection techniques},
  author={Scalabrino, Simone and Bavota, Gabriele and Linares-V{\'a}squez, Mario and Piantadosi, Valentina and Lanza, Michele and Oliveto, Rocco},
  journal={Empirical Software Engineering},
  volume={25},
  number={6},
  pages={5006--5046},
  year={2020},
  publisher={Springer}
}

@inproceedings{pivot-icse19,
  title={Pivot: learning api-device correlations to facilitate android compatibility issue detection},
  author={Wei, Lili and Liu, Yepang and Cheung, Shing-Chi},
  booktitle={2019 IEEE/ACM 41st International Conference on Software Engineering (ICSE)},
  pages={878--888},
  year={2019},
  organization={IEEE}
}

@inproceedings{cid-issta18,
  title={Cid: Automating the detection of api-related compatibility issues in android apps},
  author={Li, Li and Bissyand{\'e}, Tegawend{\'e} F and Wang, Haoyu and Klein, Jacques},
  booktitle={Proceedings of the 27th ACM SIGSOFT International Symposium on Software Testing and Analysis},
  pages={153--163},
  year={2018}
}

@inproceedings{ictapifinder-ase18,
  title={Understanding and detecting evolution-induced compatibility issues in android apps},
  author={He, Dongjie and Li, Lian and Wang, Lei and Zheng, Hengjie and Li, Guangwei and Xue, Jingling},
  booktitle={Proceedings of the 33rd ACM/IEEE International Conference on Automated Software Engineering},
  pages={167--177},
  year={2018}
}

@inproceedings{cider-ase18,
  title={Understanding and detecting callback compatibility issues for android applications},
  author={Huang, Huaxun and Wei, Lili and Liu, Yepang and Cheung, Shing-Chi},
  booktitle={Proceedings of the 33rd ACM/IEEE International Conference on Automated Software Engineering},
  pages={532--542},
  year={2018}
}

@inproceedings{ficfinder-ase16,
  title={Taming android fragmentation: Characterizing and detecting compatibility issues for android apps},
  author={Wei, Lili and Liu, Yepang and Cheung, Shing-Chi},
  booktitle={Proceedings of the 31st IEEE/ACM International Conference on Automated Software Engineering},
  pages={226--237},
  year={2016}
}

@article{ficfinder-tse2018,
  title={Understanding and detecting fragmentation-induced compatibility issues for android apps},
  author={Wei, Lili and Liu, Yepang and Cheung, Shing-Chi and Huang, Huaxun and Lu, Xuan and Liu, Xuanzhe},
  journal={IEEE Transactions on Software Engineering},
  volume={46},
  number={11},
  pages={1176--1199},
  year={2018},
  publisher={IEEE}
}

@inproceedings{psdroid-icse2023,
  title={Compatibility Issue Detection for Android Apps Based on Path-Sensitive Semantic Analysis},
  author={Yang, Sen and Chen, Sen and Fan, Lingling and Xu, Sihan and Hui, Zhanwei and Huang, Song},
  booktitle={45th International Conference on Software Engineering},
  year={2023}
}

@inproceedings{rapid-icse2020,
  title={How Android developers handle evolution-induced API compatibility issues: a large-scale study},
  author={Xia, Hao and Zhang, Yuan and Zhou, Yingtian and Chen, Xiaoting and Wang, Yang and Zhang, Xiangyu and Cui, Shuaishuai and Hong, Geng and Zhang, Xiaohan and Yang, Min and others},
  booktitle={Proceedings of the ACM/IEEE 42nd International Conference on Software Engineering},
  pages={886--898},
  year={2020}
}

@inproceedings{mcdonnell-icsm2013,
  title={An empirical study of api stability and adoption in the android ecosystem},
  author={McDonnell, Tyler and Ray, Baishakhi and Kim, Miryung},
  booktitle={2013 IEEE International Conference on Software Maintenance},
  pages={70--79},
  year={2013},
  organization={IEEE}
}

@inproceedings{aper-icse2022,
  title={Aper: evolution-aware runtime permission misuse detection for Android apps},
  author={Wang, Sinan and Wang, Yibo and Zhan, Xian and Wang, Ying and Liu, Yepang and Luo, Xiapu and Cheung, Shing-Chi},
  booktitle={Proceedings of the 44th International Conference on Software Engineering},
  pages={125--137},
  year={2022}
}

@inproceedings{yang-icse2022,
  title={Demystifying Android non-SDK APIs: measurement and understanding},
  author={Yang, Shishuai and Li, Rui and Chen, Jiongyi and Diao, Wenrui and Guo, Shanqing},
  booktitle={Proceedings of the 44th International Conference on Software Engineering},
  pages={647--658},
  year={2022}
}

@inproceedings{confdroid-ase21,
  title={Characterizing and detecting configuration compatibility issues in android apps},
  author={Huang, Huaxun and Wen, Ming and Wei, Lili and Liu, Yepang and Cheung, Shing-Chi},
  booktitle={2021 36th IEEE/ACM International Conference on Automated Software Engineering (ASE)},
  pages={517--528},
  year={2021},
  organization={IEEE}
}

@inproceedings{junittestgen-ase22,
  title={Mining android api usage to generate unit test cases for pinpointing compatibility issues},
  author={Sun, Xiaoyu and Chen, Xiao and Zhao, Yanjie and Liu, Pei and Grundy, John and Li, Li},
  booktitle={37th IEEE/ACM International Conference on Automated Software Engineering},
  pages={1--13},
  year={2022}
}

@inproceedings{mimic-icse19,
  title={Mimic: UI compatibility testing system for Android apps},
  author={Ki, Taeyeon and Park, Chang Min and Dantu, Karthik and Ko, Steven Y and Ziarek, Lukasz},
  booktitle={2019 IEEE/ACM 41st International Conference on Software Engineering (ICSE)},
  pages={246--256},
  year={2019},
  organization={IEEE}
}

@inproceedings{conffix-issta23,
  title={ConfFix: Repairing Configuration Compatibility Issues in Android Apps},
  author={Huang, Huaxun and Xu, Chi and Wen, Ming and Liu, Yepang and Cheung, Shing-Chi},
  booktitle={ACM SIGSOFT International Symposium on Software Testing and Analysis},
  year={2023}
}

@inproceedings{repairdroid-icse22,
  title={Towards automatically repairing compatibility issues in published android apps},
  author={Zhao, Yanjie and Li, Li and Liu, Kui and Grundy, John},
  booktitle={Proceedings of the 44th International Conference on Software Engineering},
  pages={2142--2153},
  year={2022}
}

@inproceedings{cda-msr18,
  title={Characterising deprecated android apis},
  author={Li, Li and Gao, Jun and Bissyand{\'e}, Tegawend{\'e} F and Ma, Lei and Xia, Xin and Klein, Jacques},
  booktitle={Proceedings of the 15th International Conference on Mining Software Repositories},
  pages={254--264},
  year={2018}
}

@inproceedings{li-icsme16,
  title={Accessing inaccessible android apis: An empirical study},
  author={Li, Li and Bissyand{\'e}, Tegawend{\'e} F and Le Traon, Yves and Klein, Jacques},
  booktitle={2016 IEEE International Conference on Software Maintenance and Evolution (ICSME)},
  pages={411--422},
  year={2016},
  organization={IEEE}
}

@inproceedings{linares-fse13,
  title={Api change and fault proneness: A threat to the success of android apps},
  author={Linares-V{\'a}squez, Mario and Bavota, Gabriele and Bernal-C{\'a}rdenas, Carlos and Di Penta, Massimiliano and Oliveto, Rocco and Poshyvanyk, Denys},
  booktitle={Proceedings of the 2013 9th joint meeting on foundations of software engineering},
  pages={477--487},
  year={2013}
}

@inproceedings{linares-icpc14,
  title={How do api changes trigger stack overflow discussions? a study on the android sdk},
  author={Linares-V{\'a}squez, Mario and Bavota, Gabriele and Di Penta, Massimiliano and Oliveto, Rocco and Poshyvanyk, Denys},
  booktitle={Proceedings of the 22nd International Conference on Program Comprehension},
  pages={83--94},
  year={2014}
}

@inproceedings{liu-icseseip21,
  title={Identifying and characterizing silently-evolved methods in the android API},
  author={Liu, Pei and Li, Li and Yan, Yichun and Fazzini, Mattia and Grundy, John},
  booktitle={2021 IEEE/ACM 43rd International Conference on Software Engineering: Software Engineering in Practice (ICSE-SEIP)},
  pages={308--317},
  year={2021},
  organization={IEEE}
}

@inproceedings{cai-issta19,
  title={A large-scale study of application incompatibilities in Android},
  author={Cai, Haipeng  and  Zhang, Ziyi  and  Li, Li  and  Fu, Xiaoqin},
  booktitle={the 28th ACM SIGSOFT International Symposium on Software Testing and Analysis},
  year={2019}
}

@inproceedings{appevolve-issta19,
  title={Automated API-usage update for Android apps},
  author={Fazzini, Mattia and Xin, Qi and Orso, Alessandro},
  booktitle={Proceedings of the 28th ACM SIGSOFT international symposium on software testing and analysis},
  pages={204--215},
  year={2019}
}

@article{bavota2014impact,
  title={The impact of api change-and fault-proneness on the user ratings of android apps},
  author={Bavota, Gabriele and Linares-Vasquez, Mario and Bernal-Cardenas, Carlos Eduardo and Di Penta, Massimiliano and Oliveto, Rocco and Poshyvanyk, Denys},
  journal={IEEE Transactions on Software Engineering},
  volume={41},
  number={4},
  pages={384--407},
  year={2014},
  publisher={IEEE}
}

@inproceedings{liu2022automatically,
  title={Automatically detecting api-induced compatibility issues in android apps: a comparative analysis (replicability study)},
  author={Liu, Pei and Zhao, Yanjie and Cai, Haipeng and Fazzini, Mattia and Grundy, John and Li, Li},
  booktitle={Proceedings of the 31st ACM SIGSOFT International Symposium on Software Testing and Analysis},
  pages={617--628},
  year={2022}
}

@inproceedings{saintdroid-dsn22,
  title={SAINTDroid: Scalable, Automated Incompatibility Detection for Android},
  author={Silva, Bruno and Stevens, Clay and Mansoor, Niloofar and Srisa-An, Witawas and Yu, Tingting and Bagheri, Hamid},
  booktitle={2022 52nd Annual IEEE/IFIP International Conference on Dependable Systems and Networks (DSN)},
  pages={567--579},
  year={2022},
  organization={IEEE}
}

@inproceedings{mahmoudi2018android,
  title={The android update problem: An empirical study},
  author={Mahmoudi, Mehran and Nadi, Sarah},
  booktitle={Proceedings of the 15th International Conference on Mining Software Repositories},
  pages={220--230},
  year={2018}
}

@article{li2019understanding,
  title={Understanding and analyzing java reflection},
  author={Li, Yue and Tan, Tian and Xue, Jingling},
  journal={ACM Transactions on Software Engineering and Methodology (TOSEM)},
  volume={28},
  number={2},
  pages={1--50},
  year={2019},
  publisher={ACM New York, NY, USA}
}

@inproceedings{lam2011soot,
  author={Lam, Patrick and Bodden, Eric and Lhot{\'a}k, Ondrej and Hendren, Laurie},
  title={The Soot Framework for Java Program Analysis: A Retrospective},
  booktitle={Cetus Users and Compiler Infastructure Workshop (CETUS 2011)},
  year={2011}
}

@inproceedings{wang2021understanding,
  title={Understanding and facilitating the co-evolution of production and test code},
  author={Wang, Sinan and Wen, Ming and Liu, Yepang and Wang, Ying and Wu, Rongxin},
  booktitle={2021 IEEE International conference on software analysis, evolution and reengineering (SANER)},
  pages={272--283},
  year={2021},
  organization={IEEE}
}

@inproceedings{cao2022rotten,
  title={Rotten apples spoil the bunch: An anatomy of Google Play malware},
  author={Cao, Michael and Ahmed, Khaled and Rubin, Julia},
  booktitle={Proceedings of the 44th International Conference on Software Engineering},
  pages={1919--1931},
  year={2022}
}

@article{tam2017evolution,
  title={The evolution of android malware and android analysis techniques},
  author={Tam, Kimberly and Feizollah, Ali and Anuar, Nor Badrul and Salleh, Rosli and Cavallaro, Lorenzo},
  journal={ACM Computing Surveys (CSUR)},
  volume={49},
  number={4},
  pages={1--41},
  year={2017},
  publisher={ACM New York, NY, USA}
}

@article{martin2016survey,
  title={A survey of app store analysis for software engineering},
  author={Martin, William and Sarro, Federica and Jia, Yue and Zhang, Yuanyuan and Harman, Mark},
  journal={IEEE transactions on software engineering},
  volume={43},
  number={9},
  pages={817--847},
  year={2016},
  publisher={IEEE}
}

@inproceedings{el2021dissecting,
  title={Dissecting residual APIs in custom android ROMs},
  author={El-Rewini, Zeinab and Aafer, Yousra},
  booktitle={Proceedings of the 2021 ACM SIGSAC Conference on Computer and Communications Security},
  pages={1598--1611},
  year={2021}
}

@inproceedings{dong2018understanding,
  title={Understanding android obfuscation techniques: A large-scale investigation in the wild},
  author={Dong, Shuaike and Li, Menghao and Diao, Wenrui and Liu, Xiangyu and Liu, Jian and Li, Zhou and Xu, Fenghao and Chen, Kai and Wang, Xiaofeng and Zhang, Kehuan},
  booktitle={Security and Privacy in Communication Networks: 14th International Conference, SecureComm 2018, Singapore, Singapore, August 8-10, 2018, Proceedings, Part I},
  pages={172--192},
  year={2018},
  organization={Springer}
}

@article{a3-tse2020,
  title={A3: Assisting android api migrations using code examples},
  author={Lamothe, Maxime and Shang, Weiyi and Chen, Tse-Hsun Peter},
  journal={IEEE Transactions on Software Engineering},
  volume={48},
  number={2},
  pages={417--431},
  year={2020},
  publisher={IEEE}
}

@article{apimatchmaker-tse22,
  title={APIMatchmaker: Matching the Right APIs for Supporting the Development of Android Apps},
  author={Zhao, Yanjie and Li, Li and Wang, Haoyu and He, Qiang and Grundy, John},
  journal={IEEE Transactions on Software Engineering},
  volume={49},
  number={1},
  pages={113--130},
  year={2022},
  publisher={IEEE}
}

@article{acid-tse23,
  title={Detecting Android API Compatibility Issues With API Differences},
  author={Mahmud, Tarek and Che, Meiru and Yang, Guowei},
  journal={IEEE Transactions on Software Engineering},
  year={2023},
  publisher={IEEE}
}

@misc{web:android/androidx,
  title        = {AndroidX Overview | JetPack | Android Developers},
  url          = {https://developer.android.com/jetpack/androidx},
  lastaccessed = {May 29, 2023},
year={2023}
}

@misc{web:android/lint,
  title        = {Improve your code with lint checks},
  year         = {2018},
  url          = {https://developer.android.com/studio/write/lint},
  lastaccessed = {May 25, 2023}
}

@misc{web:android/non-sdk-api,
  title        = {Restrictions on non-SDK interfaces},
  year         = {2018},
  url          = {https://developer.android.com/guide/app-compatibility/restrictions-non-sdk-interfaces},
  lastaccessed = {May 25, 2023}
}

@misc{web:android/veridex,
  title        = {Test using the veridex tool},
  year         = {2018},
  url          = {https://developer.android.com/guide/app-compatibility/restrictions-non-sdk-interfaces#test-veridex-tool},
  lastaccessed = {May 25, 2023}
}

@misc{web:android/api-diff-report,
  title        = {Android API Differences Report},
  url          = {https://developer.android.com/sdk/api_diff/33/changes},
  lastaccessed = {May 29, 2023},
  year = {2023}
}

@misc{web:android/sdk-manager,
  title        = {sdkmanger},
  url          = {https://developer.android.com/tools/sdkmanager},
  lastaccessed = {June 8, 2023},
  year = {2023}
}

@misc{web:android/current-txt,
  title        = {current.txt},
  url          = {https://android.googlesource.com/platform/frameworks/base/+/refs/heads/main/core/api/current.txt},
  lastaccessed = {June 12, 2023},
  year         = {2024}
}

@misc{web:stackoverflow/update-api,
  title        = {How does android generate 'current.txt' it's api description file?},
  year         = {2017},
  url          = {https://stackoverflow.com/questions/42036298/},
  lastaccessed = {June 5, 2023}
}

@misc{web:java/jvm2-2.9,
  title        = {The Structure of the Java Virtual Machine},
  year         = {2014},
  url          = {https://docs.oracle.com/javase/specs/jvms/se8/html/jvms-2.html#jvms-2.9},
  lastaccessed = {June 6, 2023}
}

@misc{web:apache/android-httpclient,
  title        = {Apache HttpComponents - HttpClient for Android},
  url          = {https://hc.apache.org/httpcomponents-client-4.5.x/android.html},
  lastaccessed = {June 26, 2023},


year={2024}
}

@misc{web:khronos/egl,
  title        = {OpenGL ES Overview - The Khronos Group Inc},
  url          = {https://www.khronos.org/opengles/},
  lastaccessed = {July 5, 2023},



year={2024}
}

@misc{web:stackoverflow/load-android,
  title        = {Find all classes in a package in Android - Stack Overflow},
  url          = {https://stackoverflow.com/questions/15446036},
  lastaccessed = {July 9, 2023},
  year={2013}
}

@misc{web:samsung/getperflogenable,
  title        = {3.0 Update - Samsung Members},
  year         = {2021},
  url          = {https://r2.community.samsung.com/t5/Galaxy-M/3-0-Update/td-p/7156608},
  lastaccessed = {July 11, 2023}
}

@misc{web:github/samsung-customized-keyevent,
  title        = {pandalion98/SPay-inner-workings - GitHub},
  url          = {https://github.com/pandalion98/SPay-inner-workings/blob/master/boot.oat_files/arm64/dex/out1/android/view/KeyEvent.java#L518},
  lastaccessed = {July 11, 2023},
  year = {2017}
}

@misc{web:github/app-manager,
  title        = {MuntashirAkon/AppManager - GitHub},
  url          = {https://github.com/MuntashirAkon/AppManager},
  lastaccessed = {July 27, 2023},
  year={2024}
}

\end{document}